# Fast Current Regulation and Persistent Current Maintenance of High-Temperature Superconducting Magnets with Contact Power Supply and Flux Pump

Chenghuai Wu(吴成怀), Wei Wang (王为)[1)], Run Long(龙润), Hong Li(李洪), Li Zhou(周立), Peng Liu(刘鹏)

College of Electrical Engineering, Sichuan University, Chengdu 610065, China

*Abstract*—Due to the properties of high temperature superconducting (HTS) materials, current attenuation is inevitable during the closed-loop operation of HTS magnets. When a contact DC power supply is used to supplement this attenuation, it inevitably creates a huge thermal burden on the cryogenic system. The flux pump is a revolutionary new power source that can charge closed-loop HTS magnet wirelessly. However, for HTS magnets with a large inductance, such as particle accelerator magnets and magnetic confinement magnet in Tokamak devices, the flux pump cannot fast adjust the DC current of the magnet, due to its small DC output voltage. Here, we present a method to fast regulate the current in a closed-loop HTS magnet using a contact DC power supply and persistent current switch (PCS). After current regulation, the HTS magnet is operated in the persistent current mode (PCM) with a flux pump. By applying the "four-quadrant" control theory of the flux pump allows, the current in HTS magnet is controlled with high stability. This study provide a power strategy for the fast current regulation and maintenance of persistent current in the HTS magnet, enabling the industrial applications of flux pumps for HTS magnets with large inductance.

*Keywords*—Flux pump, YBCO wire, persistent current mode (PCM), superconducting magnet, wireless power transfer.

## I. INTRODUCTION

THE second-generation (2G) high temperature superconducting (HTS) wire exhibit outstanding performance in a strong magnetic field, and have excellent mechanical strength and flexibility [2, 3]. In recent years, HTS magnets wound by HTS wires have made significant advancements in high-field magnets [4, 5], nuclear magnetic resonance imaging magnets [6-8], magnetic levitation [9, 10], and superconducting motors [11-13], etc. However, HTS magnets cannot operate in a closed-loop, or in the PCM like low temperature superconducting (LTS) magnets, due to the inevitable soldering resistance [14] and the low *n* value in its superconducting *E-J* power relationship [15, 16]. If the traditional contact DC power supply is used to maintain the operation current, the presence of current leads cause huge heat leakage into the cryogenic system, and its own resistance also generates extra joule heat, resulting in the extremely high energy consumption. One potential solution to this problem is to use an HTS flux pump to wirelessly inject a large amount of DC current into the closed-loop HTS magnet and operate in PCM, thus eliminating the need for current leads and contact power supply, thus decrease the energy consumption by several orders of magnitude compared with conventional contact DC power supply.

Over the last decade, a variety of HTS flux pumps have been developed to wirelessly charge HTS magnets, including HTS dynamo [17-19], linear-motor type flux pumps [20, 21], linear-pulse field flux pumps [22, 23], and transformer-rectifier flux pumps [24-28]. Of these, linear-motor type flux pump, HTS dynamo and linear-pulse field flux pumps are categorized as travelling wave flux pump [29]. Despite the prevalence of travelling wave flux pumps, the origin of their DC output has remained a theoretical challenge since their discovery. This is because it cannot be clearly explained by the induction law. Inspired by Giaever's "DC Transformer" experiment [30, 31], Wang [32] proposed macroscopic magnetic flux coupling theory to explain the source of the DC electromotive force of the travelling wave flux pump. The theory explains that the moving magnetic pole generated by the travelling wave flux pump can couple a large number of vortices on the superconducting films, and drag the vortices to move in a predetermined direction, thereby generating a DC electromotive force, given as:

$$\vec{E} = \vec{B} \times \vec{v}_f \qquad (1)$$

where $\vec{B}$ is the flux density of coupled vortices, and $\vec{v}_f$ is the velocity of the travelling magnetic pole.

Relying on Eqn.(1), Wang [1] then introduced a "four-quadrant" control method to accurately control the DC output of HTS travelling wave flux pumps. By controlling the direction of the travelling magnetic wave or the DC bias magnetic field, accurate control of the pumped current in the magnet is achieved, making the method promising for applications where high current accuracy is crucial. For example, in a reported 14 T MRI, 1400 A of current need to be passed to a superconducting magnet with a 300 H inductance, and the current ripple cannot exceed 1 ppm [33, 34]. In this study, we controlled the switch of the DC bias coil of the flux

*Supported by the National Natural Science Foundation of China under grant numbers 51877143, and the Science and Technology Project of Sichuan Province, China under grant number 2021YFS0088.

1) E-mail: weiwangca283@gmail.com

pump via feedback, allowing to maintain PCM at any value in both directions below the maximum output current value of the flux pump. The current ripple is less than 5‰, making it suitable for most applications involving superconducting magnets. These results validates the viability of the "four-quadrant" control theory for achieving high-precision control of current ripple in travelling wave flux pumps.

On the other hand, compared with conventional contact DC power supplies, the travelling wave flux pumps have relatively low output voltage [35], which have relatively slow current regulation rate for the HTS magnet with large inductance. However, in some application scenarios, fast excitation [36] and demagnetization [37, 38] or even fast current regulation [39-41] is necessary. For instance, in the cyclotron magnet, the superconducting magnet is required to be excited smoothly to 4-6 T within 1-10 s, and in the Tokamak device, superconducting magnets with caliber of 10-20 m are required to be excited to the maximum field intensity of 3-10 T within 1-20 s. To cope with this problem, here we propose a method to fast adjust the current in an HTS magnet operated in the PCM. During the current regulation stage, the contact DC power supply and PCS are used for fast current regulations. During the closed-loop operation phase, the flux pump is used to supplement the current attenuation caused by the soldering resistance. This method combines the advantages of contact DC power supplies and flux pumps, which can realize both the fast current regulation and PCM operation for HTS magnets. The research results of this work can provide a novel power strategy to enable the use of flux pumps for HTS magnets with large inductances, thus decrease the energy consumption by several orders of magnitude, which may accelerate the broad application of HTS magnets.

## II. WORKING PRINCIPLE

### A. Process and circuit principles for fast current regulation and maintain PCM operation

The power strategy proposed in this work enables the fast switching between two operation modes: fast current regulation and the PCM operation. The equivalent circuit is shown in Fig. 1, which contains two circuit loops, the first is the current regulation loop composed of a contact DC power supply $I_s$ and HTS coil $L_{DPC}$, the second is the PCM operation loop comprised of the flux pump, HTS bridge (stator) and the HTS coil $L_{DPC}$. The switching between the two operation modes are controlled by two switches $S_1$ and $S_2$, while $S_2$ is the PCS controlled by a heater, i.e., when the heater is switched on, the HTS bridge becomes normal state, the PCS is "open". $R_{j1}$ and $R_{j2}$ are soldering resistances, $R_c$ is the resistance of the current leads, and $L_{DPC}$ is the inductance of the HTS DPC. $I_L$ is the current through DPC and $I_2$ is the current through the bridge when switch $S_2$ is closed.

$V_{dc}$ and $R_d$ [42] are DC e.m.f. and internal resistance generated by the flux pump, $S_3$ and $S_4$ are switches of the DC bias coil and AC coil of the flux pump respectively, i.e., based on Eqn.(2) and (3), by switching on the DC bias coil and the AC coil of the flux pump, which is equivalent to $S_3$ and $S_4$ is

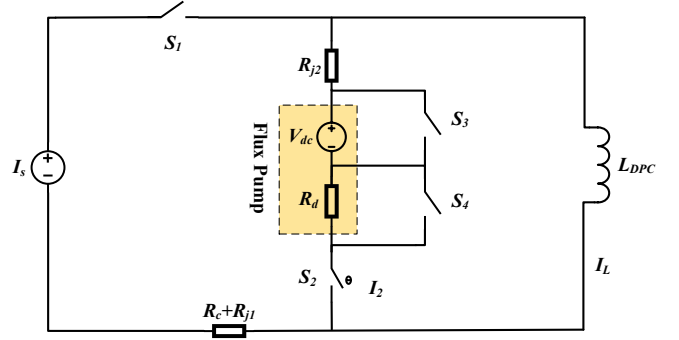

Fig. 1. Equivalent circuit diagram of the fast current regulation and PCM operation of an HTS magnet.

"open", the DC e.m.f. $V_{dc}$ is switched on, and current is pumped into the closed-loop. It is worth noting that only when the DC bias coil of the flux pump and the AC coil are opened at the same time, the flux pump will generate DC e.m.f. in the closed loop, only open the DC bias coil will not generate DC e.m.f..

**1) Fast excitation and maintain PCM operation.**

Fig. 2 shows the operation procedure and circuit principle of fast excitation and maintain PCM operation. First, turn on the heater (open the switch $S_2$), turn on the contact power supply (close the switch $S_1$), and adjust the current of contact DC power supply $I_s$ so that $I_L$ can reach its ideal value. The reason for this operation is to use the contact power supply for fast excitation of the DPC. Second, after the fast excitation is completed, it is necessary to switch to the closed-loop operation mode. Turn off the heater (close the switch $S_2$) and slowly reduce the contact DC power supply $I_s$ current to 0 A. In this process, $I_s + I_2 = I_L$, where $I_L$ remains unchanged due to the inductance of DPC, $I_S$ decreases to 0 A, $I_2$ becomes the same as $I_L$, and DPC enters closed-loop operation. Finally, disconnect the contact power supply (open the switch $S_1$) and turn on the flux pump power supply (open the switch $S_3$ and $S_4$) to maintain PCM operation.

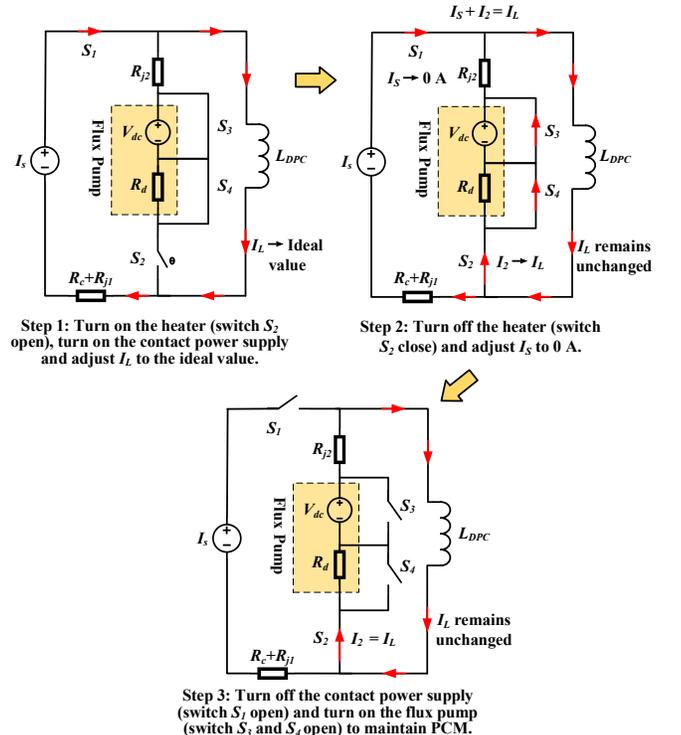

Fig. 2. The operation procedure and circuit principle of fast excitation and maintain PCM operation.

**2) Fast demagnetization.**

Fig. 3 shows the operation procedure and circuit principle of fast demagnetization. First of all, ensure that the closed-loop operation magnet is in the natural decay state. If the flux pump is turned on, it needs to be turned off first (keep switch $S_3$ and $S_4$ close). Secondly, turn on the contact power supply (close the switch $S_1$) and adjust the knob until the current $I_2$ is 0 A. In this process, $I_S + I_2 = I_L$, where $I_L$ remains unchanged due to the inductance of DPC, $I_2$ decreases to 0 A, $I_S$ becomes the same as $I_L$, and DPC enters open loop operation. Finally, turn on the heater (open the switch $S_2$) and reduce the current of the contact power supply to 0 A. So that $I_L = I_S = 0\,A$, fast demagnetization is complete.

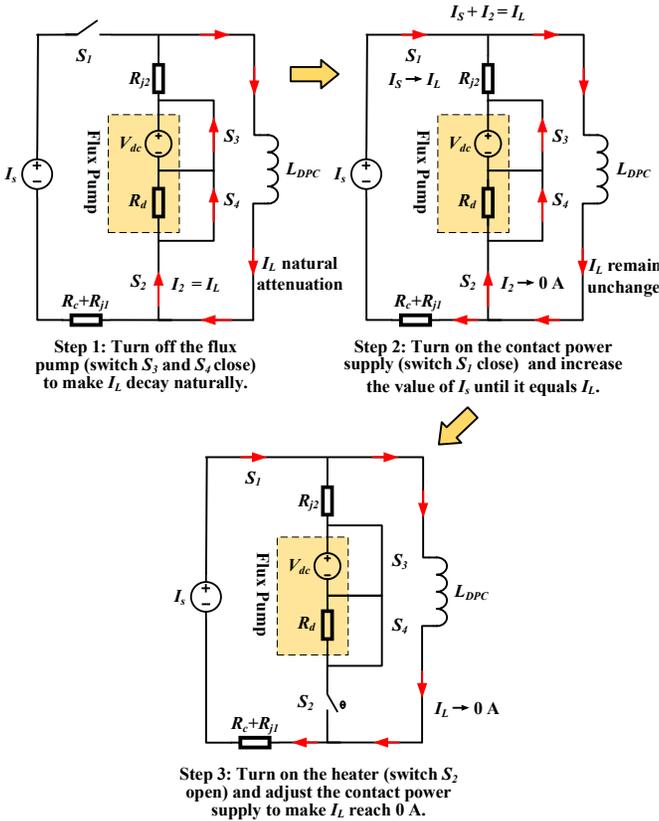

Fig. 3. The operation procedure and circuit principle of fast demagnetization.

**3) Fast adjust the current and maintain PCM operation.**

Fig. 4 shows the operation procedure and circuit principle of fast adjust the current and maintain PCM operation. First of all, ensure that the closed-loop operation magnet is in the natural decay state. If the flux pump is turned on, it needs to be turned off first (keep switch $S_3$ and $S_4$ close). Secondly, turn on the contact power supply (close the switch $S_1$) and adjust the knob until the current $I_2$ is 0 A. In this process, $I_S + I_2 = I_L$, where $I_L$ remains unchanged due to the inductance of DPC, $I_2$ decreases to 0 A, $I_S$ becomes the same as $I_L$, and DPC enters open loop operation. Third, turn on the heater (open the switch $S_2$) and adjust the current of the contact power supply to the ideal value. In this process, $I_L = I_S$, completing the fast regulation of the current in the DPC. Fourth, turn off the heater (close the switch $S_2$) and slowly reduce the contact DC power supply $I_S$ current to 0 A. In this process, $I_S + I_2 = I_L$, where $I_L$ remains unchanged due to the inductance of DPC, $I_S$ decreases to 0 A, $I_2$ becomes the same as $I_L$, and DPC enters closed-loop operation. Finally, disconnect the contact power supply (open the switch $S_1$) and turn on the flux pump power supply (open the switch $S_3$ and $S_4$) to maintain PCM operation.

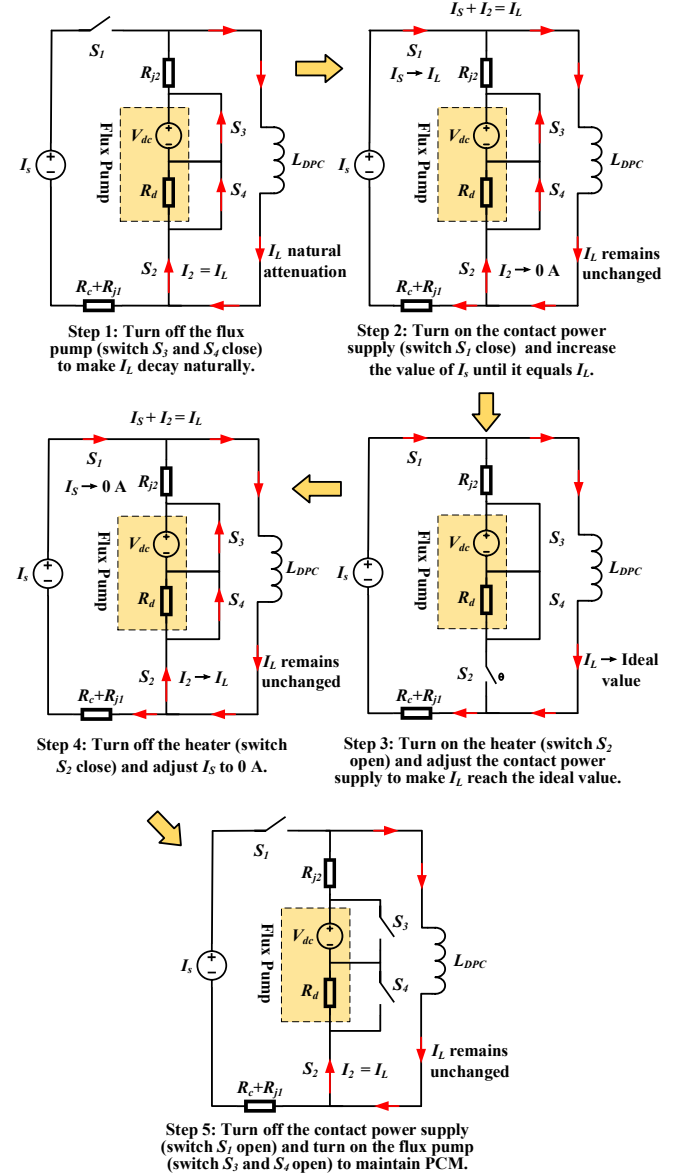

Fig. 4. The operation procedure and circuit principle of fast adjust the current and maintain PCM operation.

### B. Working principle of linear-motor type flux pump

The linear-motor type flux pump is mainly composed of AC coils and DC bias coil, which generates a DC biased AC travelling magnetic wave in the air gap, as shown in Fig. 5, which can be expressed by a one-dimensional wave equation [43] as:

$$B_y(x,t) = B_{ac} \sin(kx + \omega t) + B_{dc} \quad (2)$$

where $B_{ac}$ is the amplitude of AC travelling magnetic wave, $B_{dc}$ is the value of DC bias field, $k = \frac{2\pi}{\lambda}$ is the wave number, $\lambda$ is the wavelength, $\omega = 2\pi f$ is the angular frequency, and $f$ is the frequency.

The DC biased AC travelling magnetic wave generated by the travelling wave flux pump can couple a large number of superconducting vortices on the superconducting stator, and drag the vortices to move in a predetermined direction. According to Eq. (1), the vortices move in a predetermined direction results in an output DC e.m.f., expressed as:

$$V_{dc} = l_{eff} B_y \times v_x \quad (3)$$

where $V_{dc}$ is the averaged DC output e.m.f., $l_{eff}$ is the effective length along the YBCO stator, $B_y$ is the average flux density of coupled vortices, and $v_x$ is the velocity of the travelling magnetic pole.

### C. "Four-quadrant" control theory accurately controls DC output

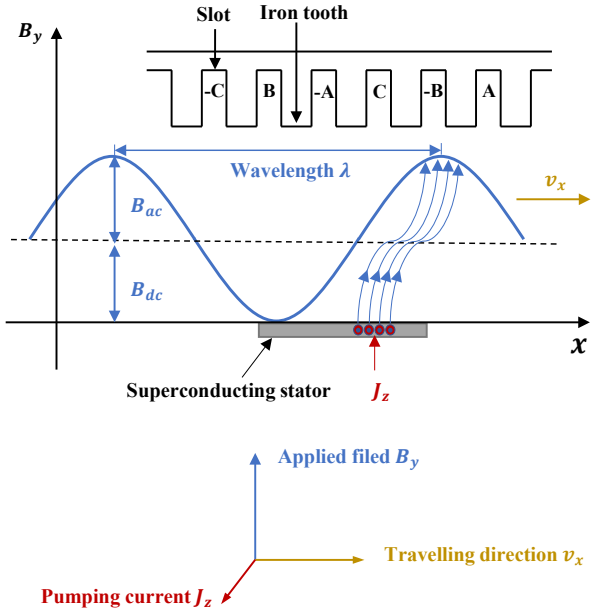

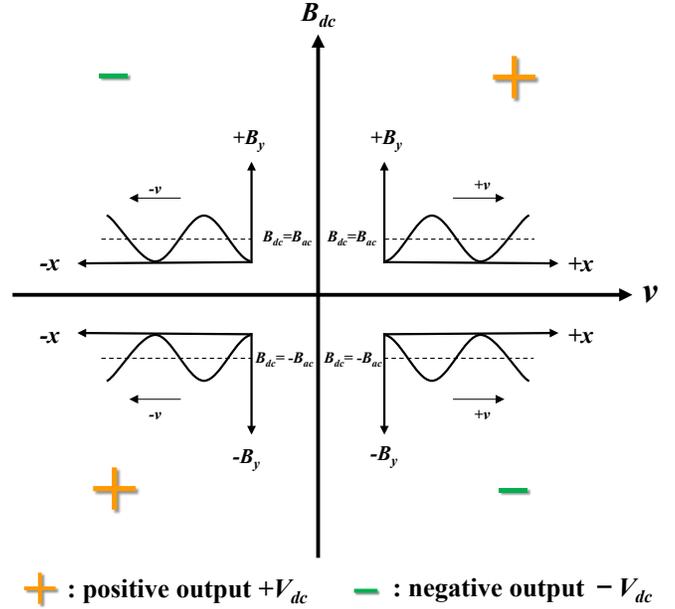

Fig. 6. Schematic diagram of "four-quadrant" control. This control method proposed by Wang [1] to accurately control the DC output of HTS travelling wave flux pumps, based on the theory of macroscopic magnetic coupling effect.

Fig. 5. The applied magnetic field $B_y$ is perpendicular to the superconducting stator and the pumped current is perpendicular to the 2D plane.

Relying on Eqn.(2) and Eqn.(3), a "four-quadrant" method was introduced by Wang [1] to accurately control the DC output of HTS travelling wave flux pumps, as demonstrated in Fig. 6, i.e., by reversing either the direction of DC bias field $B_{dc}$ or the travelling direction $v_x$, the DC output e.m.f. $V_{dc}$ can be reversed in the HTS travelling wave flux pumps. Based on the "four-quadrant" control method, we designed a feedback control algorithm to control the on/off of the DC bias coils of the flux pump. we first run the flux pump at its maximum output capacity, i.e., ensuring $|B_{dc}| = B_{ac}$ [43]. Then a feedback control program is used to control the DC bias coils of the flux pump to ensure that the pump current can be output at the preset current $I_{preset}$ within an allowed fluctuation $\Delta I$, which take any value and stabilize it as shown in Fig. 7. To accomplish the target, the feedback program will track the difference between the output current and the preset current in real time, i.e., when the output current value is less than the preset value, the DC bias coils is turned on for charging, and when the current value is greater than the preset value, the DC bias coils is turned off for attenuation.

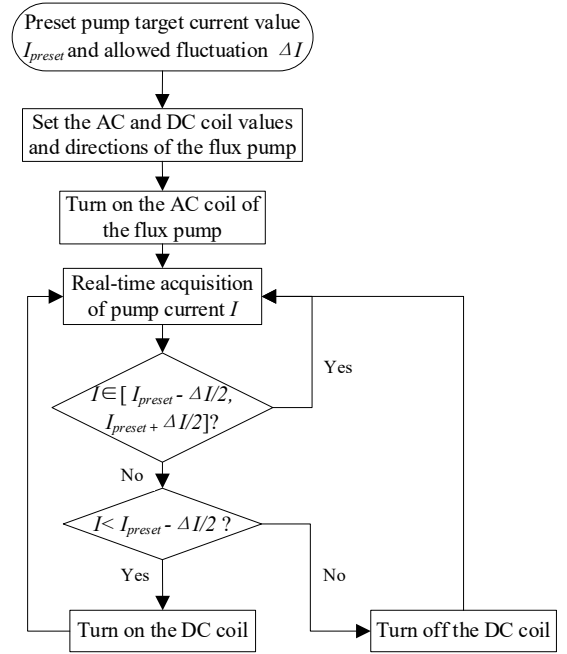

Fig. 7. Block diagram of feedback control. By comparing the difference between the pump current and the preset current, we perform a real-time feedback control of the pump current.

### III. EXPERIMENTAL

#### A. Experimental setup

In the experiments, the entire closed-loop HTS coil shown in Fig. 8 is submerged in liquid nitrogen at 77 K. A contact DC power supply is connected to the terminals of the HTS coil by two current leads, as shown in Fig. 8, to fast excite or demagnetize the HTS coil. After fast current regulation, the flux pump is used to maintain PCM operation of the superconducting magnet, and the heater is used to switch on/off of the PCS. The 12 mm wide YBCO wire made by *Shanghai Superconductor Technology Company* [44] is used as the stator in the air gap of the flux pump. An HTS closed-loop is formed by connecting the DPC and the superconducting stator by soldering, while the measured soldering resistance is 30 $n\Omega$. In addition, two Hall sensors are installed in the middle of the DPC and the C-shaped iron ring to measure the magnetic field. By calculating the measured magnetic field, the current in both the DPC and the stator can be obtained. Detailed parameters of HTS DPC, stator and heater are shown in Table I.

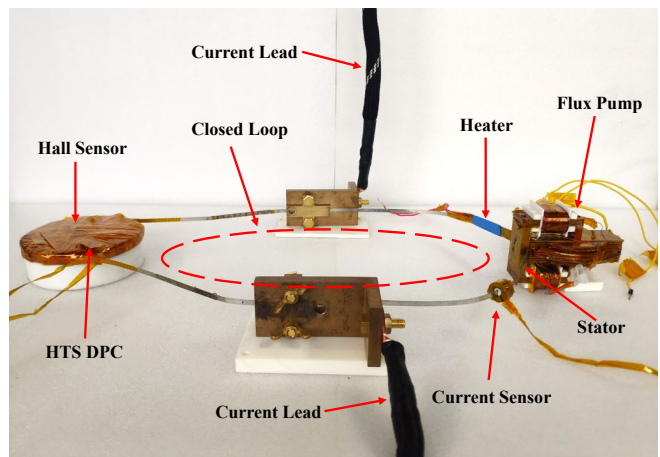

Fig. 8. The connection of the experimental apparatus. The current leads are used to fast adjust the current in the DPC, the heater is used as PCS and the flux pump is used to maintain the HTS magnet's PCM operation.

TABLE I
MAIN SPECIFICATIONS OF EXPERIMENTAL SETUP

| Parameters | Specification |
|---|---|
| HTS INS DPC | |
| Material / Manufacturer | YBCO / SSTC |
| Wire Width / Length | 4.8 mm / 50 m |
| inner diameter / Outer diameter | 50 mm / 125 mm |
| Number of turns | 88×2=176 turns |
| Inductance | 2.52 mH |
| $I_c$@77k, Self-field | 82 A |
| Stator | |
| Material / Manufacturer | YBCO / SSTC |
| Thickness / Width | 220 μm / 12 mm |
| Welding resistance | 30 $n\Omega$ |
| $I_c$@77k, Self-field | 325 A |
| Heater | |
| Size | 10×93×0.25 mm |
| Heater resistance | 12 $\Omega$ |
| Rated temperature | 443 K |

### B. Power Supplies, Measurement and Control System

The AC coils of the linear-motor type flux pump are power by a three-phase inverter. The DC bias coils of the linear-motor type flux pump are powered by a programmable DC power supply. The current leads are connected to a contact superconducting DC power supply (Keysight 6680A), which can output a maximum current of 875 A. The two Hall sensors, installed in the center of the DPC and C-ring to measure current, are powered by a precise DC current supply.

The data acquisition system is mainly composed of data acquisition instrument (Agilent 34972A), National Instruments PCI-4070 cards, which read the voltages from the two Hall sensors and calibrate to the value of pumped current. To accurately control the closed-loop current, the measured current data is transmitted to a LabView feedback control program. Based on the algorithm shown in Fig. 7, the LabView program controls the on/off state of the DC power supply connected to the DC bias coil of the flux pump.

## IV. RESULT AND DISCUSSION

### A. Fast excitation and maintain PCM operation

The conventional contact DC power supply has the advantage of fast excitation speed, while the flux pump has the advantage of maintaining PCM operation. We combine the advantages of the two power supplies and propose a method of "fast regulation with contact DC power supply and maintain PCM operation with flux pump". In order to visually reflect the effect of the flux pump in maintaining PCM operation, we used the contact DC power supply to fast excitation the HTS magnet, then allow current attenuation in the closed-loop and use the flux pump to compensate the current losses, which enables the PCM of the closed-loop HTS magnet. The experimental operation and circuit principle of fast excitation and maintain PCM operation are described in Section A of Chapter II. The experimental results are shown in Fig. 9. In the experiment where only contact power was used, the current in the magnet attenuated fast after switching to the closed-loop operation due to the soldering resistances in the loop. However, in the experiment using contact power supply and flux pump, fast excitation was realized with the help of contact power supply, and after switching to flux pump to maintain PCM operation, the closed-loop current maintained constant, only a little change occurred during the switching process.

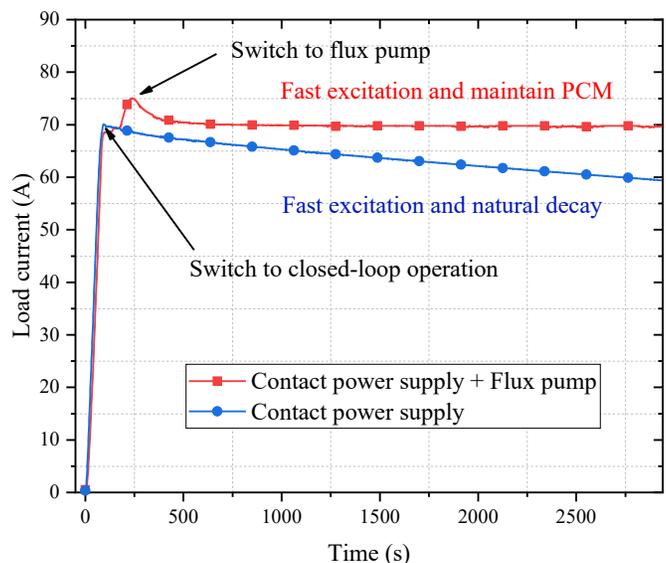

Fig. 9. Experimental results of fast excitation and maintain PCM operation experiment. The flux pump was used to maintain PCM operation experiment, no current attenuation was observed. However, the current decayed rapidly in the experiment without a flux pump to maintained PCM operation.

The conventional contact power supply can be used to excite HTS magnet rapidly, but the current lead must be pulled out because of the huge energy consumption caused by the use of contact power to maintain the current in HTS magnets. However, due to the existence of soldering resistance, the stable closed-loop operation of the HTS magnet cannot be maintained. The results of this comparison experiment show that using contact power supply to fast excitation, and then using flux pump to maintain PCM operation is an ideal solution for HTS magnets with large inductances which requires fast excitation and maintain PCM operation.

### B. Comparison of excitation and demagnetizing speeds

According to the "four-quadrant" control theory，As shown in Fig. 6，by reversing the direction of DC bias field $B_{dc}$, the DC output $e.m.f.$ $V_{dc}$ can be reversed in the HTS travelling wave flux pumps. Therefore, when the HTS magnets needs demagnetization, the current in the HTS magnets can be reverse-charged to 0 A by reversing the direction of DC bias field $B_{dc}$. Subsequently, we design experiments to compare the excitation and demagnetization speeds of the propose power strategy combining conventional contact power supply and flux pump, with the strategy of only use a flux pump. The experimental operation and circuit principles of fast excitation, maintain PCM operation, and fast demagnetization are described in Section A of Chapter II.

Fig. 10 shows the experimental results. It takes 1614 s for the flux pump to pump 69 A into the closed-loop HTS magnet. For comparison, it only takes 98 s to excite the HTS magnets to 69 A by contact DC power supply. Similarly, we compared the speed of the demagnetization speeds with the two strategies: the flux pump takes 299 s to demagnetize the current from 69 A to 0 A, while contact DC power supply takes only 68 s. The excitation speed of the contact DC power supply is more than 16 times larger than the flux pump, and the demagnetization speed is more than 4 times. In addition, the excitation and demagnetization speed of contact DC power supply can be further improved by outputting a higher voltage. Therefore, with the help of contact DC power supply, the excitation and demagnetization speed of the HTS magnets can be greatly improved.

The inductance of the HTS magnet used in this experiment is only 2.52 mH, but the excitation speed of the HTS magnet with the flux pump is 16 times slower than that of the contact power supply. If the flux pump is used to excite an HTS magnet with a large inductance, the excitation and demagnetizing speeds of flux pump is difficult to meet the industrial requirements. For instance, in the cyclotron magnet, the superconducting magnet is required to be excited smoothly to 4-6 T within 1-10 s, and in ITER magnets, fast demagnetization is required [36, 37]. The proposed combined power strategy may have the capability to solve this problem.

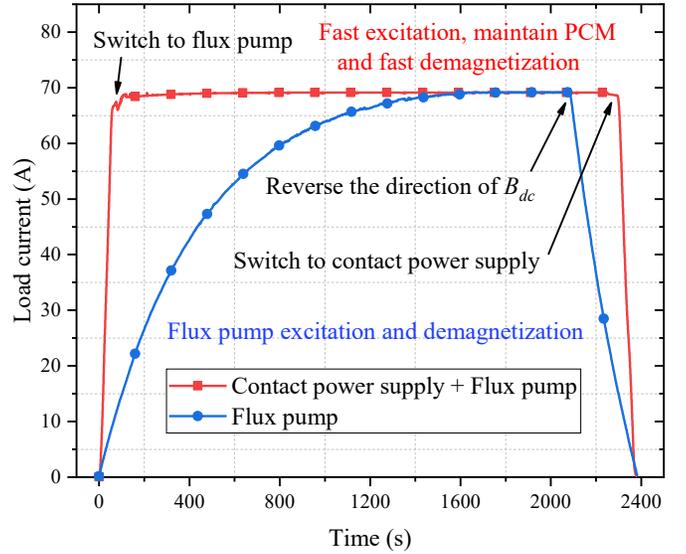

Fig. 10. Comparison of excitation and demagnetization speed between two power supply. The excitation speed of the contact power is 16 times faster than that of the flux pump, and the demagnetizing speed is 4 times faster.

### C. Fast bipolar excitation and maintain PCM operation

The above experiments demonstrate that the use of both contact DC power supply and flux pump can achieve fast excitation/demagnetization, while maintaining the PCM operation of the HTS magnet after current regulations. In applications, the load current in the HTS magnet needs to be controlled to a preset value with certain accuracy, in which case the feedback control of the flux pump's output current is required. For instance, in nuclear magnetic resonance imaging, the flux pump should not only compensate the magnetic field attenuation, but also need to minimize the ripple magnetic field to meet the required magnetic field stability.

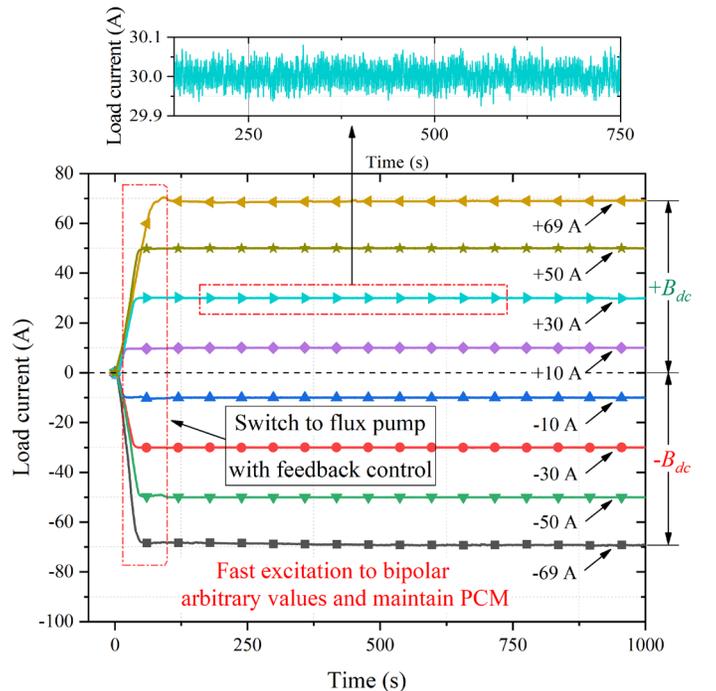

Fig.11. Experimental results of fast excitation to bipolar arbitrary value and maintain PCM operation.

Based on the "four-quadrant" control theory of travelling wave flux pump [1], control both the direction of DC bias field $B_{dc}$ and the travelling direction $v$ of AC travelling wave can control the direction of output e.m.f. of the flux pump, as shown in Fig. 6. In particular, the output current can be controlled by the on/off states of the DC bias coil or AC coils, while the control flowchart is shown in Fig. 7. In this section, the feedback current control of the flux pump based on "four-quadrant" control theory is utilized to maintain the PCM operation after fast current regulation.

The experimental results for fast charging and maintain the PCM operation are shown in Fig. 11. In the experiments, we preset the target current in the feedback control program, such as +10 A, +30 A, +50 A and +69 A, respectively, then use the contact power to fast excitation to the target value, and then switch to the flux pump and open the feedback control program to maintain the target value to maintain the persistent current. For reverse charging, the operation process is the same as above, the only difference is to reverse the direction of the DC bias filed $B_{dc}$. The procedure and circuit principle of fast excitation and maintain PCM operation are described in Section A of Chapter II.

The experimental results show that the HTS magnets can be excited fast to any preset value in both directions by using the contact power supply, and maintaining its persistent current by the flux pump based on feedback control algorithm. In addition, we checked the current fluctuation when maintained at 30 A and found that the current fluctuation was only 0.15 A and the current stability reached 5‰, which is consistent with the preset accuracy. This experiment verifies that, with the feedback current control based on the "four-quadrant" control theory, the travelling wave flux pump can work as a bipolar power supply to maintain PCM operation with high current precision.

### D. Fast current regulation and maintain PCM operation

In several applications such as the excitation coil of superconducting machine, cyclotron magnet, etc., fast regulation of the operation current is necessary. For instance, in proton heavy ion therapy, the current in the cyclotron magnet needs to be adjusted rapidly in order to kill cancer cells at different depths [41]. In this section, we demonstrate the switching from flux pump to contact power supply to enable fast current regulations, while the closed-loop current can be fast changed or even reversed in the HTS magnet.

As shown in Fig. 12, a contact power supply is used to rapidly excite the magnet, and then switch to the flux pump to maintain PCM operation. After maintaining the persistent current for a period of time, the circuit is switched back to the contact power supply to fast adjust the current to any desired value. In the experiments, we demonstrate the fast adjusted to +50 A, +30 A and +10 A, demagnetization to 0 A, and reverse adjust to -10 A, -30 A, -50 A, and -69 A. After the contact power supply fast adjusts the current, the circuit is switched to the flux pump with feedback control to maintain the persistent current in the PCM, with the current ripple less than 0.15 A. See Section A of Chapter II for the operating procedure and circuit principle of fast excitation, maintain PCM operation and fast adjust the current.

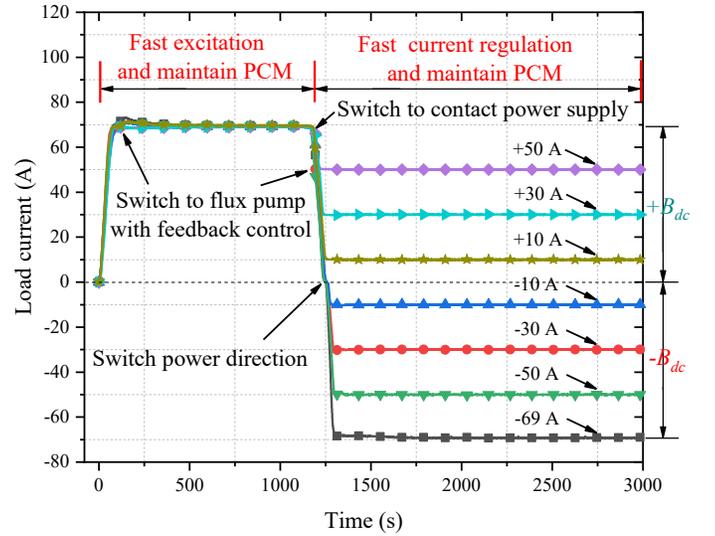

Fig. 12. Experimental results of fast adjust current to bipolar arbitrary values and maintain PCM operation.

The above experiments proves that the persistent current maintained by a flux pump in a closed-loop HTS magnet can be rapidly regulated with a contact power supply, while the PCM operation and fast current regulation mode can be rapidly switched. This power strategy incorporate both the advantages of conventional contact power and flux pump. Despite the conventional power supply has very high energy consumption, time for current regulation accounts for a very short period of time, then switched back to the ultra low energy consumption PCM operation maintained by the flux pump, while the persistent current has no current attenuation and with high accuracy.

### E. Discussion

HTS magnets are used in many fields, such as accelerator physics, Tokamak devices, proton heavy ion therapy, nuclear magnetic resonance imaging, and maglev trains, etc. The operation currents in these magnets are ranging from hundreds of amperes to tens of thousands of amperes. If conventional contact power supplies are used to maintain the currents in the HTS magnets, the energy consumptions are extremely high, while the current leads bring heavy thermal load to the cryogenic system.

Fig. 13 demonstrates the two different operation modes, such as with the contact power supply and with the flux pump, respectively. As shown in Fig. 13(a), the PCM operation maintained by contact power supply has welding resistance $R_j$ and current lead resistance $R_c$. The energy consumption required by contact power supply to maintain PCM operation is $W_c = I^2(R_{c1} + R_{c2} + R_j)$. As shown in Fig. 13 (b), the PCM operation maintained by the flux pump has welding resistance $R_j$ and internal resistance $R_d$ of flux pump in the superconducting loop due to the action of the travelling magnetic wave field on the superconducting wire [42], so the energy consumption of the flux pump during operation is $W_{fp} = I^2(R_d + R_J)$. When a thousand or ten thousand amps of current is required, the resistance of the current lead is generally 4-6 orders of magnitude larger than the internal resistance of the flux pump. As a result, the energy consumption of the two power sources to maintain PCM operation differs by 4-6 orders of magnitude. And that doesn't take into account the extra heat that the current leads introduce. In practice, the current leads will also bring additional heat burden to the refrigeration equipment as they pass through the low and high temperature environments, as shown in Fig. 13 (a). The thermal conductivity of copper at 293 K is 397 W/m.K. In addition, the thermal conductivity power is related to the internal and external temperature difference and contact area. Superconducting magnets in Tokamak devices often require tens of thousands of amperes of current, the size of the current lead is very huge, and the heat leakage caused by it cannot be ignored. Therefore, the advantage of flux pump to maintain PCM operation is highlighted, which can maintain the closed-loop operation of high-current magnet without attenuation with very low energy consumption.

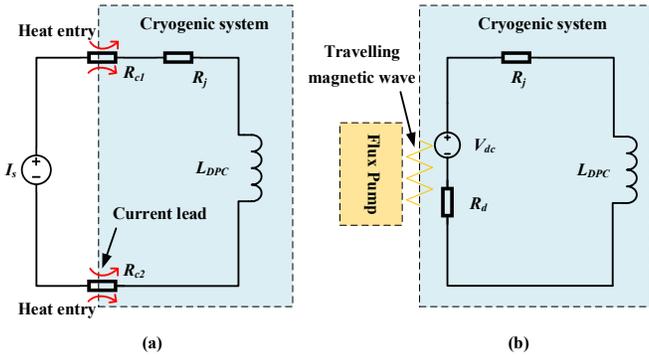

Fig. 13. Two ways to maintain PCM operation. (a) Using contact power supply to maintain PCM operation, there are problems of heat leakage and large energy consumption of current lead resistance. (b) PCM is maintained by a flux pump, with only internal resistance and welding resistance causing energy consumption.

In addition, based on the "four-quadrant" control theory, we controlled the on/off of the DC bias coil of the flux pump by feedback, and realized that the flux pump could maintain PCM operation at any value of positive or negative, with a ripple of 0.15 A and a current stability of 5‰, as shown in Fig. 11. When the flux pump is applied to the medical magnetic resonance system, the research direction is to make efforts to minimize the magnetic field ripple while compensating the magnetic field attenuation to meet the magnetic resonance system's requirements for magnetic field stability [33]. At present, because the output current is controlled only by switching on/off the DC bias coil of the flux pump, the AC coil is always on, which causes the superconducting wire to generate internal resistance under the action of the alternating magnetic field, and there will be large current ripple and decay of persistent current. Next, we plan to achieve accurate current control by controlling the on/off of the AC travelling magnetic wave field, which will further improve the current stability and further reduce the energy consumption of the flux pump. This method has great application prospect in the fields that require high current stability such as nuclear magnetic resonance.

Although the flux pump combined with feedback control has the ability to output current at positive or negative values and maintain PCM operation with high accuracy and stability, due to the low e.m.f. generated by the flux pump, the speed will be very slow if the industrial magnets with large inductances are charged by the flux pump. As shown in Fig. 10, for the magnet whose inductance is 2.52 mH used in this test, the excitation and demagnetization speeds of the flux pump are only 1/16 and 1/4 of the contact power supply, which cannot meet the industrial requirements to excite the magnet to the target magnetic field in a short period of time.

In order to solve this problem, this paper proposes the use of contact power supply in excitation, demagnetization and current regulation phase, and the use of flux pump power supply in the PCM operation maintenance phase. And when the contact power supply for current regulation, the speed of regulation is proportional to the voltage that the contact power supply can provide. After achieving the target current, the flux pump is used to maintain persistent current, which significantly reduces the energy consumption compared with continue to use the contact power supply to sustain the load current. Pluggable current leads can also be used to further decrease the heat leakage into the cryogenic system during PCM operation.

As shown in Fig. 12, the current in the magnet can be fast adjusted by switching between the closed-loop operation mode of the magnet to the open-loop operation mode, which is very promising in many application scenarios. For instance, in the high gradient magnetic separation device, the excitation speed generally reaches 2 T/min, and the maximum working field intensity is 5 T, and the stable magnetic field needs to be maintained for a long time under different intensities [40]. If this regulation method can be used, it will be of great significance to reduce operating costs and improve working efficiency.

The successful experiment of this operation mode combines the advantages of both the contact power supply and the flux pump power supply, which can not only fast adjust the current of superconducting magnets, but also maintain PCM operation in an ultra low power consumption. It can meet the requirements of fast excitation, demagnetization, current regulation and maintain PCM operation of industrial HTS magnets with large inductances.

## V. Conclusion

This paper proposes the use of a flux pump to maintain PCM operation in HTS magnets, which are used in fields such as accelerator physics, Tokamak devices, proton heavy ion therapy, nuclear magnetic resonance imaging, and maglev trains. The use of flux pump greatly reduces the energy consumption of PCM operation, because the resistance and heat conduction of the current lead of the conventional contact DC power supply will bring huge heat burden to the cryogenic system. The on/off of the DC bias coil of the flux pump is controlled by feedback, so that the flux pump can maintain the PCM of the magnet in preset positive or negative value with high stability. The current stability is 5‰ and the current ripple is as low as 0.15 A, which can meet the industrial application of most HTS magnets.

In addition, for industrial magnets with large inductance, such as particle accelerator magnets and magnetic confinement magnet in Tokamak devices, the speed of flux pump for current regulation is slow, so it is proposed to use contact DC power supply for fast excitation, fast demagnetization and fast current regulation, while flux pump is used to maintain the PCM operation after reaching the target current. This new operating mode combines the advantages of the contact DC power supply and the flux pump power supply, and can meet the requirements of fast current regulation and PCM operation maintenance of industrial HTS magnets with large inductance. This research results can provide reference for fast current regulation and maintain PCM operation of HTS magnets and accelerate the wide application of HTS magnets and travelling wave flux pumps.